# PLASMA DIAGNOSTIC AND PERFORMANCE OF A PERMANENT MAGNET HALL THRUSTER

*José Leonardo Ferreira, João Henrique Campos de Souza, Israel da Silveira Rêgo and*

*Ivan Soares Ferreira*

*Laboratório de Plasmas, Instituto de Física, Universidade de Brasília-UnB*

*70910-900 Brasilia-DF, Brazil*

**Abstract**

Electric propulsion is now a successful method for primary propulsion of deep space long duration missions and for geosyncronous satellite attitude control. Closed Drift Plasma Thruster, so called Hall Thruster or SPT (stationary plasma thruster) was primarily conceived in USSR (the ancient Soviet Union) and since then, it has been developed by space agencies, space research institutes and industries in several countries such as France, USA, Israel, Russian Federation and Brazil. In this article, we present the main features of the Permanent Magnet Hall Thruster (PMHT) developed at the Plasma laboratory of the University of Brasilia. The idea of using an array of permanent magnets, instead of an electromagnet, to produce a radial magnetic field inside the cylindrical plasma drift channel of the thruster is very significant, specially because of the possibility of developing a Hall Thruster with power consumption low enough to be used in small and medium size satellites. Descriptions such as the plasma density, temperature space profiles inside and outside the thruster channel, ion temperature measurements based on Doppler broadening of spectral lines and ion energy measurements are shown for different plasma production regimes, the space plasma potential, the measured propulsion and power consumption are shown through the paper. We also compare our data with the results reached by other types of thrusters. Finally, some of the most significant events on the evolution of the electric propulsion and their contributions to the progress on the space missions are reviewed.

## I. INTRODUCTION

Space exploration is closely tied to the rocket engines development. One of the key issues necessary to accomplish long duration space missions is the jet propellant velocity and its capability of long periods of rocket acceleration. The propulsion of spacecrafts with plasmas - also called electric propulsion - was first thought by Robert Goddart in USA (1912) and independently followed by Herman Oberth in Germany (1930) [1]. Years later, the first technical study on the application of electric propulsion to space missions was presented on an article by Ernest Stuhlinger at the 1955 International Astronautical Congress in Viena. Since then, many authors - assisted by several plasma physics methods and techniques - have been examining the necessary requirements for the construction of propulsion systems based on charged particles accelerated by eletromagnetic fields. In fact, one of the most attractive advantages of electric propulsion is the reduction of propellant mass for a given space mission because plasma thrusters can generate higher exhaust velocities than those reached by chemical rockets based on combustion processes.

In 1964, NASA launched the first series of satellites with ion thrusters. They were based on ion source models originaly made by A. T. Forrester (1959) and developed for space purposes by H. Kaufman (1960) [2]. The best results with the Kaufman source or electron bombardment ion source was obtained on DEEP SPACE 1 mission in 2000. It was a space probe sent by JPL/NASA to study the comet Borelly. It completed 4800 hours of continuos operation with 10mN thrust.

Today, space propulsion systems based on plasmas are renewing the space mission planning because the propellant is a substantial part of the space vehicle mass. It corresponds to 55% - 65% of the mass in most of geostationary satellite missions and 70% to 80% in planetary missions. After half century of science and technology improvements, electric propulsion is now allowing significant mass reduction. In low earth orbit missions, mass reduction factors from 2 to 3 can be reached due to the higher efficiency of gas ionization and plasma charged particles acceleration method. On fig.1 the plot of propellant mass to total spacecraft ratio ($\frac{m_p}{m_0}$) is shown for several specific impulse ($I_{sp}$) values [2]. Note that the electric propulsion devices have typically $I_{sp}$ = 10000 s. Others features like the increase of maneuvering time for station keeping and easier control of satellite locations on space increases the satellite life time . Launch base effects are minimized and the window launch time is improved when plasma thrusters are used on space missions.

The Hall Thruster, also known as closed electron drift thruster or SPT (Stationary Plasma Thruster) was primarily developed in the ancient USSR (now Russia) in 1960. At that time, the simple design and working principles were the main attractive properties of this plasma thruster concept. The first space mission with a 60mN thrust SPT was made by Russia in 1972 with the Meteor satellite series [3]. After this initial sucessfull operation, more than 100 Hall Thrusters were tested in several Russian space missions. Since 1995, most of the research and development of Hall Thrusters have been made on France. A joint enterprise between CNES and SNECMA was created to develop SPT with 100 mN of total thrust. They have been operating in GTO (geostationary) telecommunication satellites since 2002. One of the most triumphant thrusters developed by the French researchers is installed in the ESA Moon mission SMART-1, launched in October 2003.

It is important to compare the closed drift thruster with others known methods of plasma propulsion. Eletrothermal or resistorjet and arcjet are based on heat mechanisms of plasma acceleration. The SPT family is part of the MPD (Magneto Plasma Dynamics) and MHD (Magneto hydrodynamic) thrusters based on eletromagnet mechanisms of acceleration. Thrusters with eletrostatic acceleration of ions are based on electron bombardment and RF ion sources with electrically polarized grid acceleration systems [4]. On fig. 2 the working regions of several plasma thrusters and chemical rockets are compared on a graph that relates the thrust ($T$) and the specific impulse ($I_{sp}$). They are given respectively by:

$$\begin{cases} T = \dfrac{dm}{dt} U \\ I_{sp} = \dfrac{U}{g} \end{cases}, \qquad (1)$$

where $U$ is the exhaust velocity of charges, $\dfrac{dm}{dt}$ is the propellant mass flow rate and $g$ is the earth gravity. Note that $U$ is directly proportional the plasma beam current $I_b$:

$$U = \dfrac{I_b}{Aqn}$$

In this equation, $A$ is the total area of the thruster acceleration channel, $q$ is the electric charge and $n$ is the charge density in the plasma beam current. Verify that the Hall thrusters have the largest working region.

In this work, we show the results of a new closed drift thruster with a one-staged magnetic layer that has been developed at the Plasma Laboratory of UnB [5]. The new feature of this thruster is a magnetic field generated by permanent ceramic magnets (ferrite). The absence of electromagnet coils is the main advantage of the Permanent Magnet Hall Thruster (PMHT). With this new arrangement it is possibe to save about 200W in energy consumption, while keeping the same thrust obtained by the SPT family. It is also possible to simplify the thruster transponder and power electronics by using the proposed arrangements of permanent magnets for Hall current generation.

**II HALL THRUSTER PRINCIPLES**

The main advantage of the SPT is its comparative low power consumption (0.1 to 1.0KW) with respect to its total thrust capability (0.1N to 10N). It allows the SPT use as primary propulsion system for long space missions, because it can produce higher plasma flux density. It is also important to point out that with the absence of polarized grids, the probability of undesired effects caused by ion sputtering decreases. The SPT can work with a single cathod and requires low number of components to be controlled which means easier access and maintenance of the plasma source components.

A schematic drawing of the Hall Thruster principles is shown on fig. 3. It consists of a dieletric channel, a ring anode, an external cathod and a radial magnetic field. The basic operation principle is described by the condutivity tensor for the Hall Thruster, given by:

$$S = \begin{pmatrix} \sigma_0 & 0 & 0 \\ 0 & -\sigma_H & \sigma_H \\ 0 & \sigma_\perp & \sigma_\perp \end{pmatrix}. \qquad (2)$$

The Hall conductivity component of the matrix ($\sigma_H$) are associated with the Hall current $\vec{J}$ and with the radial magnetic field $\vec{B}$. Thus the acceleration of a plasma fluid element is given by:

$$\frac{d\vec{U}}{dt} = \frac{\vec{J} \times \vec{B}}{n} \qquad (3)$$

In the proper operating regime, the electron Hall conductivity parameter is larger than unity and the electron Larmor radius ($R_L \approx 4mm$) has to be small compared to the typical channel depth $L$ ($L \approx 38mm$), in order to magnetize only the electrons. On the other hand, the Larmor radius of the much heavier ions ($R_{Li} \approx 100mm$) is larger than the channel dimension. For this reason, they are weakly affected by the magnetic field [6]. This condition is given by:

$$R_{Li} > L > R_{Le} \qquad (4)$$

Under the influence of the axial electric and radial magnetic fields, the electrons drift in the azimutal direction (azimuthal Hall current). Due to collisions, electrons diffuse across the magnetic field towards the anode (axial electron current) and ionize by impacts the working gas atoms emerging form the anode. The axial dependence of the electron mobility depends on the magnetic field profile. This is the main physical mechanism to determine the axial space plasma potencial drop associated with the exausted plasma beam current.

### III PLASMA SOURCE AND DIAGNOSTICS DESCRIPTION

A standard glass bell jar vacuum system with a volume of $0.2m^3$ (see fig.4a) fore pump with pump velocity of $35m^3$/h and diffusion pump velocity of $500l/s$ is able to maintain $10^{-4}$torr to $10^{-5}$torr of working pressure in the vacuum chamber. In order to clean the system between working periods, a cryogenic trap is used to decrease the bell jar background pressure to $10^{-6}$ torr.

The Hall thruster plasma source is positioned at the bottom of the bell jar in order to allow a plasma drift length of 0.7m. The plasma plume near the channel (fig.4b), is clearly similar to the SPT but it is also possible to see a plasma beam going away from the plasma source indicating that some plasma is accelerated by $\vec{E} \times \vec{B}$ forces in the PMHT within an extended acceleration zone outside the channel [7]. The plasma source chamber is made of stainless steel with ionization channel cover by a thin ceramic layer within 2mm thickness. The anode ring is 2cm wide and 1mm thick. It is also made of stainless steel and it is positioned 3.8cm from the exit of the channel. Behind the anode ring, the propellant gas is uniformly distributed in the Hall plasma source chamber by using an isolated copper circular tube with several small holes.

The Hall thruster schematics of the plasma source and circuits can be seen on fig. 5. It shows a cross section view of the thruster with permanent magnet locations, anode ring, thermionic cathode, gas feed tube, electrical supplies and plasma diagnostics. The plasma is generated by a discharge between the anode ring and a negatively polarized directed heated cathode. This simple thermionic cathode is made of a 5.5cm long and 2mm thick tungsten wire cover by BaO in order to increase electron emission. By using this type of cathode it is possible to move it and choose its best position outside the ionization channel, which is 3.0cm outside the channel exit.

Electric and magnetic fields geometry are key issues in the permanent magnet Hall thruster design. The shape of these fields controls ion trajectories and the acceleration zone. Although a desirable magnetic field can be obtained by adjusting permanent magnet relative positions around the channel, a computer simulation study of possible magnetic field geometries was performed using FEMM (Finite Element Method Magnetics). The best magnetic field design was chosen from these studies.

Fig. 6 shows computer simulated magnetic field lines of the Hall thruster for two concentric circles with 76 permanent magnets bars (16 on the internal cylinder) with 6.0cm length and cross section of 1.0cm x 2.0cm. The radial component of the magnetic field for several distances from the source axis is shown on fig. 7. The magnetic field space profile along the channel has a maximum value of 350 gauss, near the channel exit and 0.5cm from the external wall. At the center of the channel a magnetic field gradient $\frac{dB}{dz} > 0$ was obtained with this magnet arrangement. It is important to point out that in this arrangement, other magnetic field components (axial and poloidal) are also present in the thruster channel. They may contribute to the generation of better plasma conditions in the ionization channel and higher plasma drift flux and the Hall current in the plasma source channel [8].

Plasma diagnostics inside and outside the source were made by using a movable Langmuir probe and an ion energy analyzer. The cylindrical Langmuir probe (0.25mm x 3.5mm) can be oriented parallel or perpendicular to the thruster axis in order to provide plasma density, plasma potencial and temperature space profiles. The probe size was designed to minimize pertubation on the plasma. We obtained the plasma density, temperature and distribution function using the Druyvensteyn method, which consists on taking the second derivative of the characteristic curve *(I x U)* measured with the Langmuir probe [10]. The distribution function *F* can be determined from the retarding potential of the probe (*U*):

$$F(U) = \frac{\sqrt{8m}}{e^{\frac{3}{2}} A} \frac{d^2 I_e}{dV_{ap}^2}$$

Once we determine *F(U)* (Fig. 16), the plasma density and the effective electron temperature are given by:

$$n_e = \int_0^\infty F(U)\sqrt{U}\, dU \quad ; \quad \langle T_e \rangle = \frac{2}{3}\int_0^\infty F(U) U^{\frac{3}{2}} dU$$

Willing to measure the plasma beam energy profile, we used a plane moveable electrode known as the ion "Faraday cup", which allows the measurement of the drifting plasma flux current space distribution and the correspondent ion energy. The ion probe was positioned 35cm from the acceleration channel exit. It was fixed with an experimental arrangement that allowed radial and angular movements permitting the sketch of the space distribution profile of the accelerated plasma .The obtained data are showed in fig.13, where we can see some asymmetry that occurs because of the use of a hot filament cathode just above the channel exit. In fig.13, the relative cathode angular position is -10° from the vertical line.

In order to perform a non-perturbative measurement of the ion temperature inside the channel, we use a plasma spectroscopy technique. It is based on the well known line broadening of spectral lines caused by several atomic processes inside the plasma.

Based on the analysis of the experimental conditions in the plasma Hall current channel, we were able to determine the appropriate statistical model to be applied on the interpretation of the measured spectral lines. Collisional ionization and excitation, charge exchange and radioactive recombination are the predominant processes that affect the radiation emitted by the plasma, and a model which includes these phenomena is satisfactorily suitable for this experimental conditions. The broadening of spectral lines is one of the consequences of the several perturbative effects on the atomic systems caused by these processes.

Line broadening can be driven by several mechanism such as Stark and Zeeman effects when strong electric and magnetic fields are applied to the plasma. High pressures can also produce collisional line enlargement; nevertheless, inside the Hall plasma source, the Doppler Broadening of spectral lines is one of the predominant characteristics of the thermal particle motion.

In order to evaluate the ion temperature of the plasma, the Doppler broadening of the plasma radiation was measured. This method is based on the difference between the wavelength emitted by a moving ion and the wavelength that would be detected if the ion were at rest. This broadening is given by:

$$\Delta \lambda_D = \lambda_0 \frac{v_i}{c} \gamma \cos(\theta)$$

where $\gamma = \left[1 - \left(\frac{v_i}{c}\right)^2\right]^{-1}$ and $\theta$ is the angle between $v_i$ and the detector "target line".

If we consider the Maxwell velocity distribution, the measured intensity of the $\Delta \lambda_D$ broadened spectral line – which emerges as a typical gaussian distribution – is expressed by:

$$I(\lambda) = I_0 e^{-\left(\frac{\lambda}{\Delta \lambda_D}\right)^2} \quad \text{(in atomic units)}$$

In this calculation, we assume that $v_i = \left(\frac{2k_B T_i}{m_i}\right)^{\frac{1}{2}} \ll c$ and consider only the radiation emitted by the ion moving towards the detector (forward or backward). Therefore, the equation for the Doppler broadening becomes:

$$\Delta\lambda_D = 7.16\times 10^{-7}\left(\frac{T_i}{M_i}\right)^{\frac{1}{2}}, \text{ in angstroms.}$$

With this equation, we can estimate the ion temperature.

**IV EXPERIMENTAL RESULTS**

The PMHT can be easily operated with anode bias voltage going from 150Volts to 700Volts. The discharge plasma current for most of the experimental conditions ranged between 0.1 to 1.5A in argon gas pressures varying from $10^{-2}$torr to $10^{-4}$torr. A calibration curve of mass flow gas feeding to the PMHT is shown on fig. 8. In this paper, we call the 0.1A discharge the low thrust regime, and the 1.5A discharge (with 150 to 200Volts), the higher thrust operational regime. The tungsten wire covered with BaO hot cathode works with temperature of 900°C emitting 1A/cm² of primary electrons to generate the discharge.

In the highest pressure conditions, the electron plasma density and temperature space profiles inside the plasma source channel are respectively in the range of 2.0 - $2.5\times 10^{10}$part/cm³ and temperature from 30eV to 120eV (see figs. 9 and 10). Plasma potential space profile (see fig. 11) shows that inside the plasma source the anode determines the plasma potential maintaining the average anode bias voltage inside the plasma channel. Outside the channel, the potential decreases mainly due to the plasma neutralization made by the thermionically emitted electrons. It was also observed that plume expansion creates density gradient. The plasma acceleration is measured using an ion energy analyzer. The drifting ion energy distribution function is proportional to $\frac{dI}{dV}$ and a derivation of analyzer characteristic curve was made by a computational code. It clearly shows two peaks indicating that ion energies are about 350eV and 600eV, for a 650Volts discharge (see fig.12).

In spite of low frequency oscillations [9] found in the ion probe measurements, we calculated the total thrust of the PMHT. The results from the ion collector current are placed in the interval of 18mN to 39mN indicating that the PMHT is working with 27% of efficiency - as expected for this first experimental model. If the total thrust is calculated from the input power in the Hall discharge, the maximum thrust goes from 44mN to 80mN . Hence, our PMHT has only 43% of the maximum thrust that can be obtained from this discharge. These values can at least be scaled up to the thrust values of the existent SPT.

## V CONCLUSION

A new conception of closed drift Hall thruster using permanent magnets was developed. The results obtained on the magnetic field geometry and strength, the plasma density temperature, the thrust measured by ion collector probe and ion energy analyzer are motivating results to forthcoming tests.

In the ion temperature measurement, we estimated that the velocity distribution is maxwellian. However, our recent experiments showed a two-electron temperature plasma inside the acceleration channel. Even though, the Doppler broadening is a very adequate method of obtaining the ion characteristics for it is a non-intrusive technique.

In spite of the use of simple materials easily found on brazillian industrial market and the utilization of Argon instead of Xenon, the new PMHT Hall thruster allowed the production of remarkable working characteristics. The extra economy on satellite energy consumption is the main advantage of this new design.

**Figure Captions**

Fig. 1 – Simulations from rocket equation for different values of specific impulse.

Fig. 2 – Working regions for several types of plasma thrusters compared with chemical rockets.

Fig. 3 – Schematics of Hall Thruster principles with the main fields and currents components.

Fig. 4 - Schematics of the vaccum chamber and access windows (left). View of the exausted plasma plume from the PMHT (right)

Fig. 5 - Hall Thruster plasma source and circuits schematics: 1-Structure of Hall Thruster, 2-Gas input pipe, 3-Anode ring, 4-Hot filament cathod, 5-Permanent magnets and 6-Plasma Diagnostics.

Fig. 6 - Magnetic field lines mapping using a finite element computer code.

Fig. 7 - Space profiles of radial, poloidal and axial components of the magnetic field.

Fig. 8 - Mass flux of neutral gas versus working pressure.

Fig. 9 – Plasma density axial space profile. The region of maximum Hall current is specified by number (2).

Fig. 10 - Electron temperature axial space profile. The transition region of the fluid regime and ionization region are specified by numbers, (3) and (4), respectively.

Fig.11 – The two-electron temperature plasma – from the Langmuir probe and the Druyvensteyn method

Fig. 12 - Plasma potential axial space profile. The acceleration region and transation region of the fluid regime are specified by numbers, (1) and (3), respectively.

Fig. 13 - Energy spectrum obtained from derivation of ion energy analyser characteristic curve. Note that the low energy peak (about 0eV) is a spurious effect of the vaccum chamber.

Fig. 14 - Angular distribution of the plasma beam.

Fig. 15 – Plasma spectrum

Fig. 16 – The Doppler broadening of the spectral lines

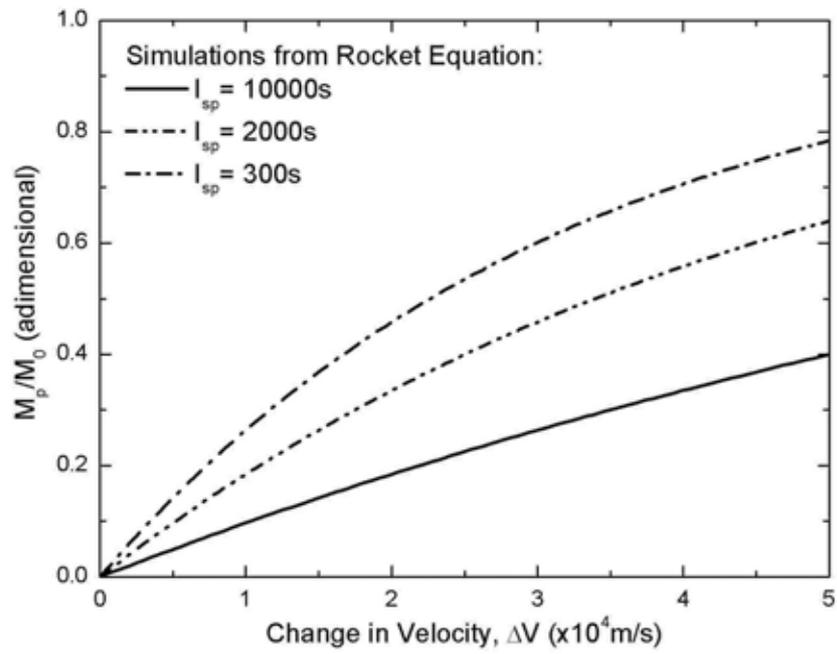

Fig.1

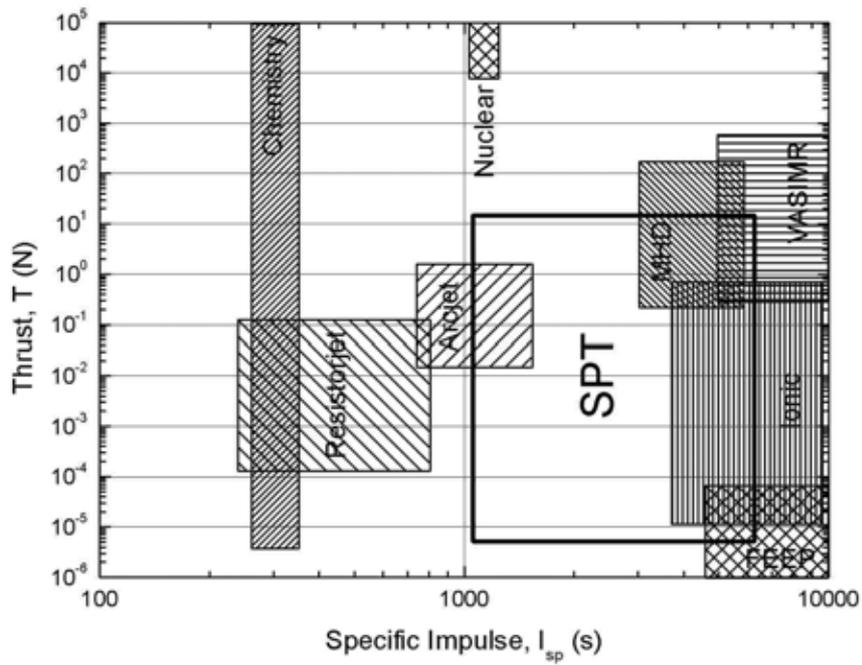

Fig. 2

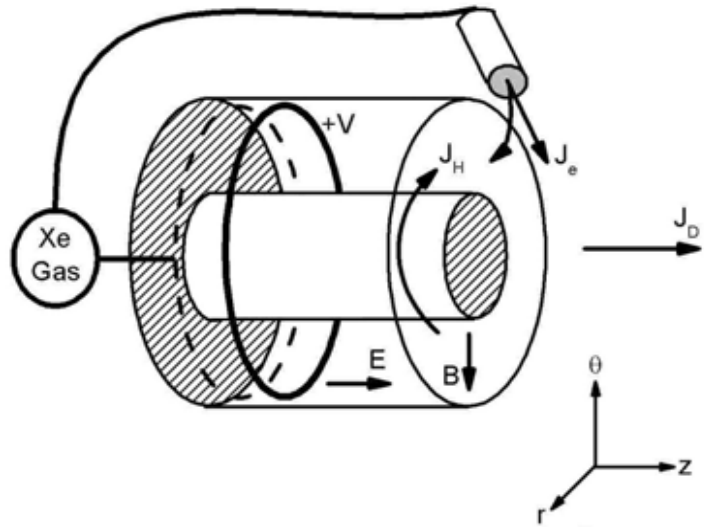

Fig. 3

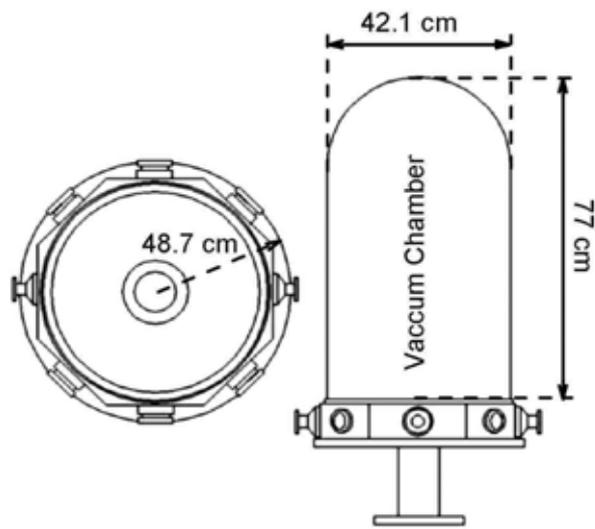

Fig. 4a

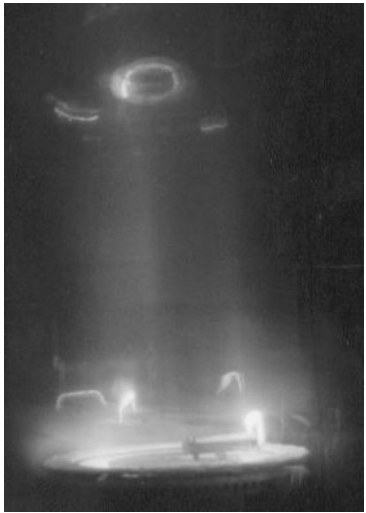

Fig. 4b

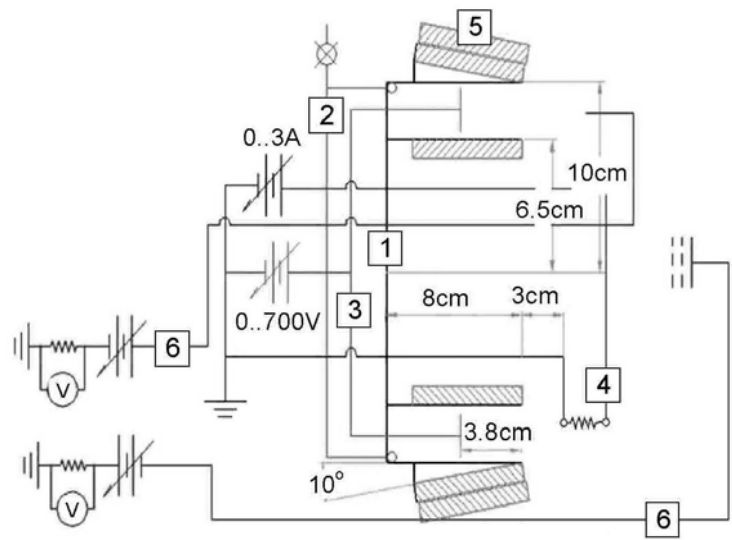

Fig. 5.

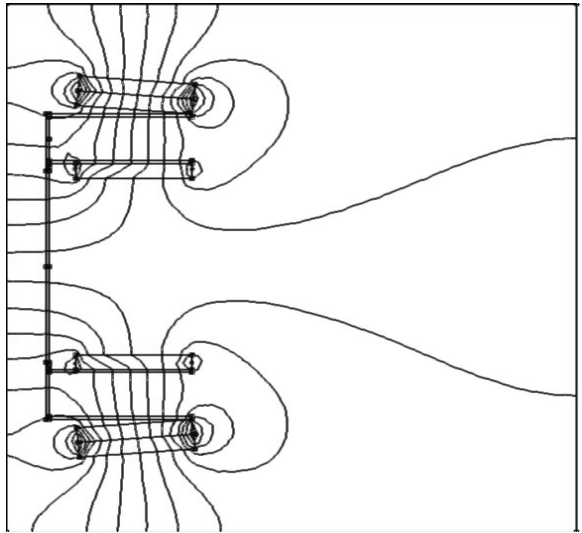

Fig. 6

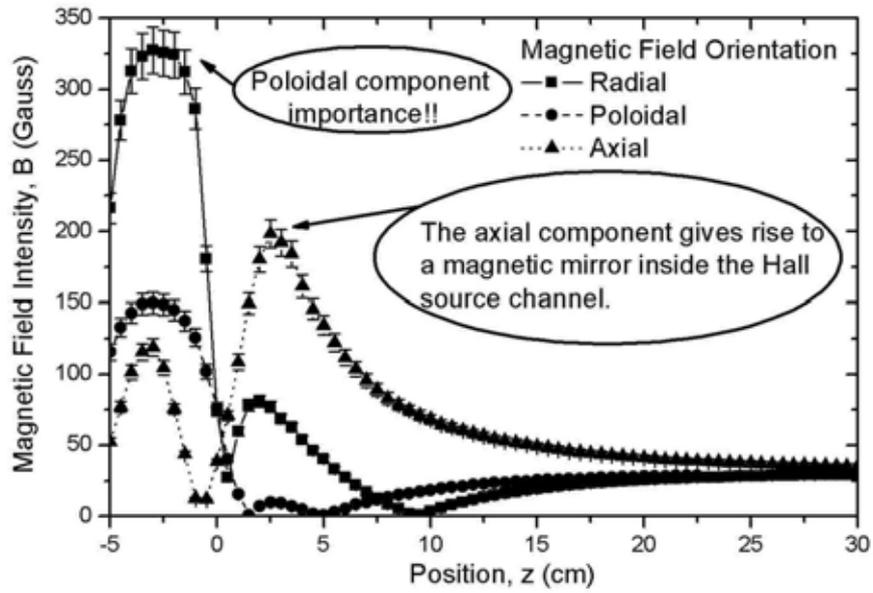

Fig. 7.

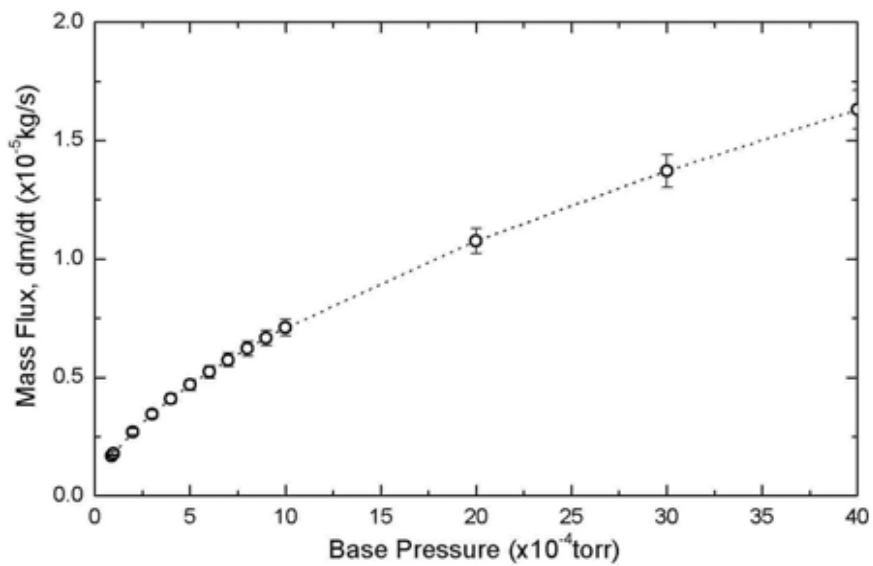

Fig. 8

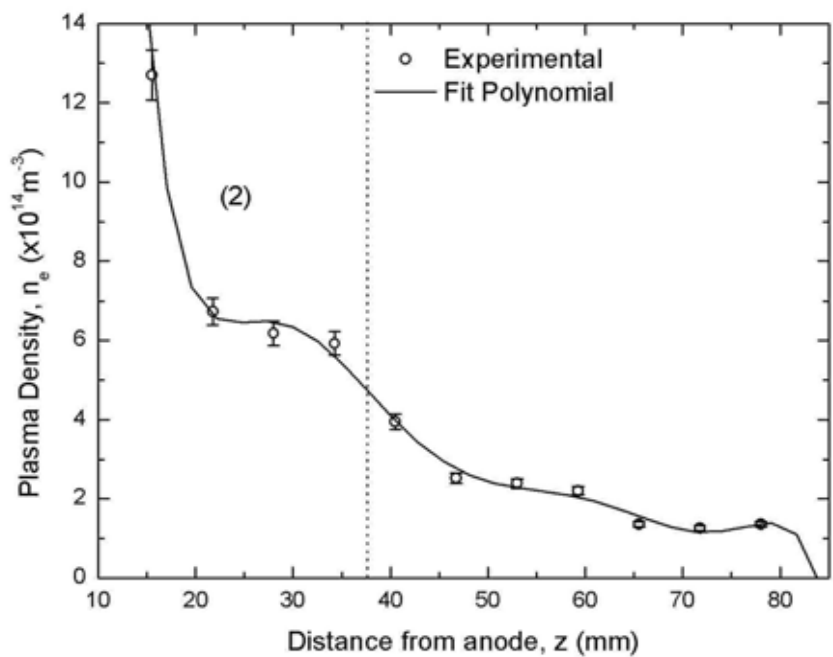

Fig. 9

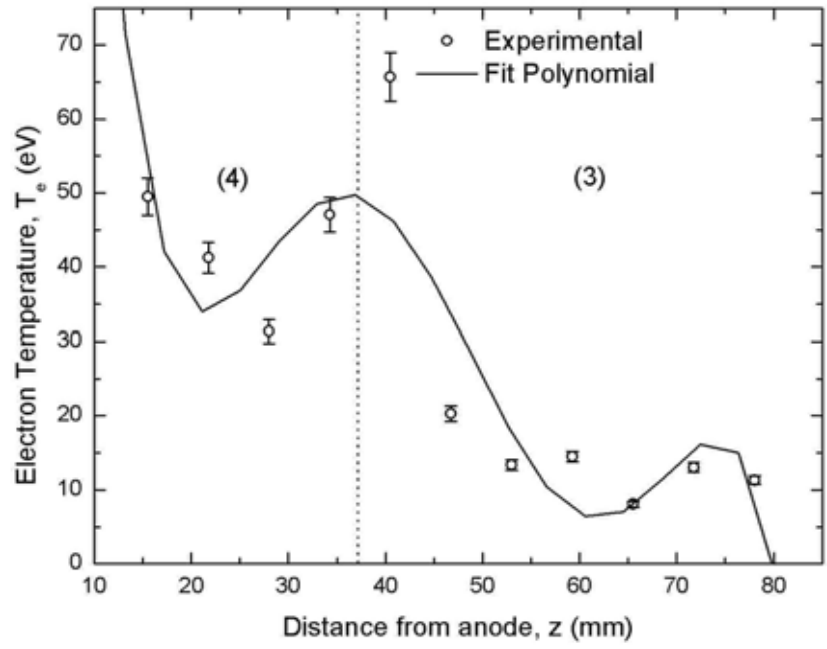

Fig.10

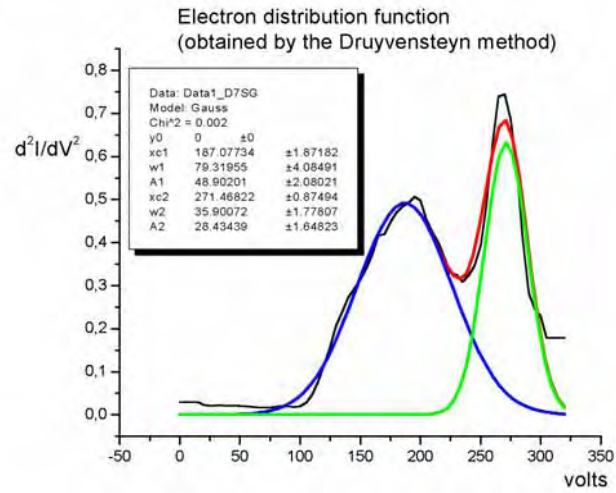

Fig.11

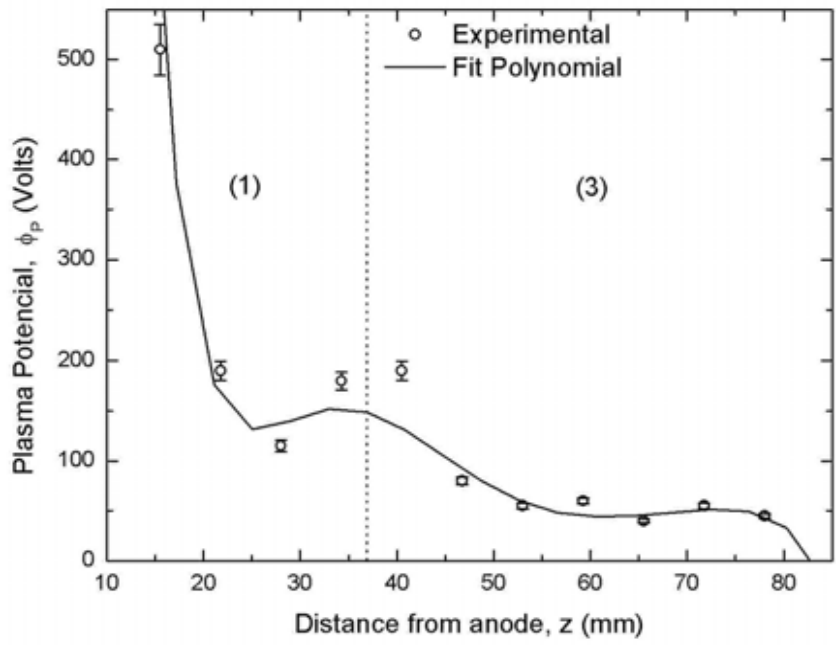

Fig.12

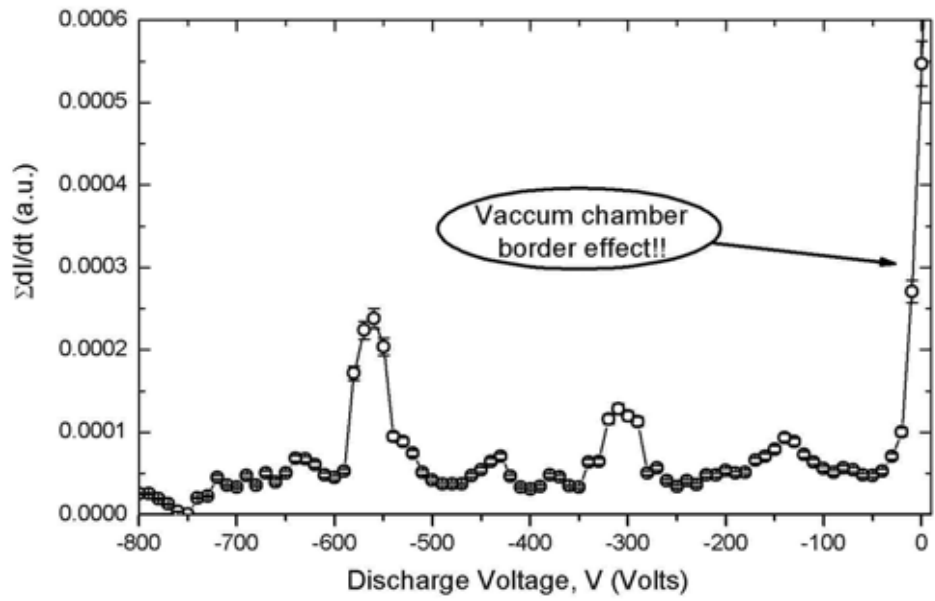

Fig.13

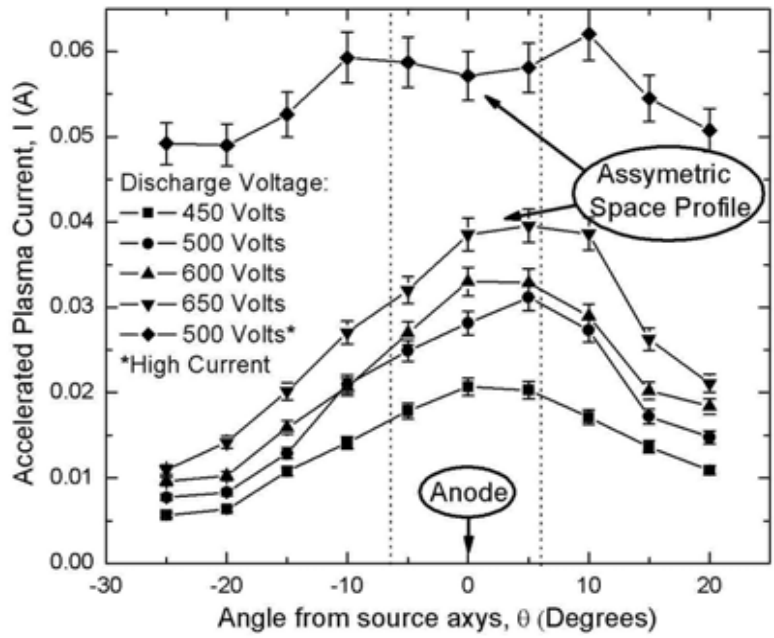

Fig.14

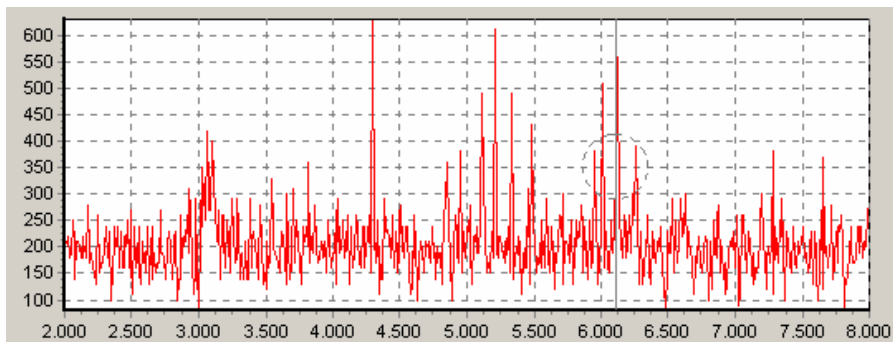

Fig. 15

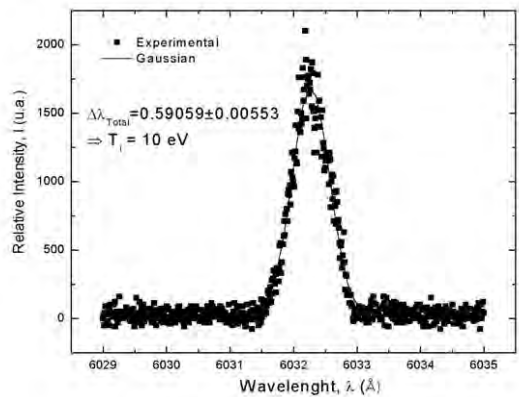

Fig. 16